\documentstyle[twocolumn,prl,aps,epsf]{revtex}
\def\Vec#1{\mbox{\boldmath $#1$}}
\begin{document}
\draft \preprint{}
\wideabs{
\title{Energy Spectrum of Superfluid Turbulence without Normal Fluid}
\author{Tsunehiko Araki$^1$, Makoto Tsubota$^1$ and Sergey K. Nemirovskii$^2$}
\address{$^1$Department of Physics,
Osaka City University, Sumiyoshi-Ku,
Osaka 558-8585, Japan \\
$^2$Institute of Thermophysics, Academy of Science, Novosibirsk 630090, Russia}
\date{Received
\hspace{25mm} } \maketitle
\begin{abstract}
The energy spectrum of the superfluid turbulence without the normal fluid is studied
numerically under the vortex filament model. Time evolution of the Taylor-Green
vortex is calculated under the full nonlocal Biot-Savart law. It is shown that for
$k<2\pi/l$, the energy spectrum is very similar to the Kolmogorov's -5/3 law which
is the most important statistical property of the conventional turbulence, where
$k$ is the wave number of the Fourier component of the velocity field and $l$ the
average intervortex spacing. The vortex length distribution becomes to obey a scaling property reflecting the self-similarity of the tangle.
\end{abstract}
 }
%
Particular attention has been focused recently on the similarity between superfluid
turbulence and conventional turbulence \cite{Vinen,Stalp,Nore}. Early work of the
superfluid turbulence has been concerned with the counterflow where the  normal fluid
 and superfluid flow oppositely \cite{Donnelly}, having no classical analog. However 
 Stalp {\it et al}. studied recently the superfluid
turbulence produced by the towed grid, thus finding the similarity between the
superfluid turbulence and the conventional turbulence above 1.4 K \cite{Stalp}. They
observed indirectly the Kolmogorov law which is one of the most important
statistical properties of the conventional turbulence. This is understood by the
idea that the superfluid and the normal fluid are likely to be coupled together by
the mutual friction between them, and to behave like a conventional fluid
\cite{Vinen,Barenghi}. Since the normal fluid is negligible at mK
temperatures, an important question arises: even free from the normal fluid, is
the superfluid turbulence still similar to the conventional turbulence or not?

As the physical model to describe the vortex dynamics in superfluid He at very low
temperatures, two types of models are well-known: the Gross-Pitaevskii (GP)
equation which describes the motion of a weakly
interacting Bose condensate, and the vortex filament model governed by
the incompressible Euler dynamics. The former reduces to the Euler vortex filament
model when variations of the wave function over scales of the order of the
superfluid healing length are neglected. The GP equation includes such complicated
compressible effects as the radiation of sound from the vortex lines
\cite{Tsubota2,Leadbeater}, the vortex-sound interactions, etc. In order to consider
the intrinsic property of superfluid turbulence in a simpler situation, we study the
energy spectrum of the 3D velocity field induced by the vortex tangle in the absence
of the normal fluid under the vortex filament model.

The energy spectrum of the vortices in superfluid was numerically calculated by
other authors. Nore {\it et al}. studied the energy spectrum of the decaying
superfluid turbulence by using the GP equation, and finding the transient spectrum
for small $k$ has the Kolmogorov law \cite{Nore}. However at late stage some
complicated compressible effects become dominant. On the other hand, an advantage of
the vortex filament model compared with the GP equation is the followings. First
this model enables us to calculate the energy spectrum free from such complicated
effects. Secondly some physical quantities, e.g. a total
line length of vortices and a vortex length distribution, can be calculated easily,
so that the relation between the statistical property in the wave number space and
the self-similarity of the vortex tangle in the real space can be discussed. Under
the vortex filament model, the energy spectra were reported on the Kelvin wave
cascade \cite{Araki,Kivotides}, which was limited to a few vortices, not a dense
tangle. Thus the calculation of the energy spectrum of a dense tangle under the
vortex filament model is expected.

For superfluid $^4$He, the vortex filament model is very useful, because the vortex
core radius $a_0 \sim 10^{-8}$cm is microscopic and the circulation $\kappa =9.97
\times 10^{-4}$cm$^2$/sec is fixed by quantum constraint. Since the Helmholtz's
theorem for a prefect fluid states that the vortex moves with the velocity produced
by themselves, the dynamics is governed by the Biot-Savart law \cite{Tsubota}. The
velocity field due to the Biot-Savart law is divided into two parts: one is the
localized induction field determined by a local curvature of vortex line, and the
other is the nonlocal field obtained by carrying out the integral of the Biot-Savart
law along the rest of the filament. When the dynamics of a dense vortex tangle is
calculated numerically, the nonlocal velocity field is usually neglected (the
localized induction approximation) \cite{Schwarz,Tsubota}. However the dynamics in
this work is calculated by the full nonlocal Biot-Savart law, because the long range
effects can be important on this problem. A vortex filament is represented by a
single string of points at a distance $\Delta \xi$ apart. When two vortices approach
within $\Delta \xi$, it is assumed that they are reconnected. The computational
sample is taken to be a cube of size $D=$ 1.0 cm. This calculation assumes the walls
of the cube to be smooth and takes account of image vortices so that the boundary
condition may be satisfied. This calculation of the dynamics is made by the
resolution $\Delta \xi =1.83 \times 10^{-2}$ cm and $\Delta t = 4.0 \times 10^{-3}$
sec.

The energy spectrum $E(k)$ is defined as $E=\int_{0}^{\infty} dk E(k)$, where E is the kinetic energy per unit mass and $k$ is the wave number of the velocity field. In our previous papers we derived the energy spectrum under the vortex filament model \cite{our papers}:
\begin{eqnarray}
E(k) & = & \frac{\kappa^2}{2(2\pi)^3}\int
\frac{d\Omega_k}{|\Vec{k}|^2}\int \int d\xi_1 d\xi_2 \nonumber \\
 &{}& \times \Vec{s}' (\xi_1)\cdot
\Vec{s}'(\xi_2) e^{-i}\Vec{^k} ^{\cdot} {^(} \Vec{^s} ^{(\xi_1)-}\Vec{^{s}}
^{(\xi_2))}, \label{eq.4}
\end{eqnarray}
where $d\Omega_k$ denotes the volume element $k^2 \sin \theta_k d \theta_k d \phi_k$
in spherical coordinates. Here a vortex filament is represented by the parametric form
$\Vec{s}=\Vec{s}(\xi,t)$, where $\Vec{s}$ refers to a point on the filament, the
prime denotes differentiation with respect to the arc length $\xi$ and the
integration is taken along the filament. The energy spectrum $E(k)$ is calculated for the vortex
configuration $\Vec{s}(\xi)$ obtained by the simulation of the dynamics.

Figure 1 shows the decay of the vortex tangle without the mutual friction
\cite{movie}. As the initial configuration of vortices, we use the Taylor-Green
vortex \cite{Nore} (Fig. 1 (a)). These initial vortices are highly
polarized. However through the chaotic dynamics which includes lots of
reconnections, the vortices become a homogeneous and isotropic vortex tangle (Fig. 1
(b), (c) and (d)).

\begin{figure}[tbhp]
\begin{minipage}{1.0\linewidth}
\begin{center}
\epsfxsize=8cm \epsfbox{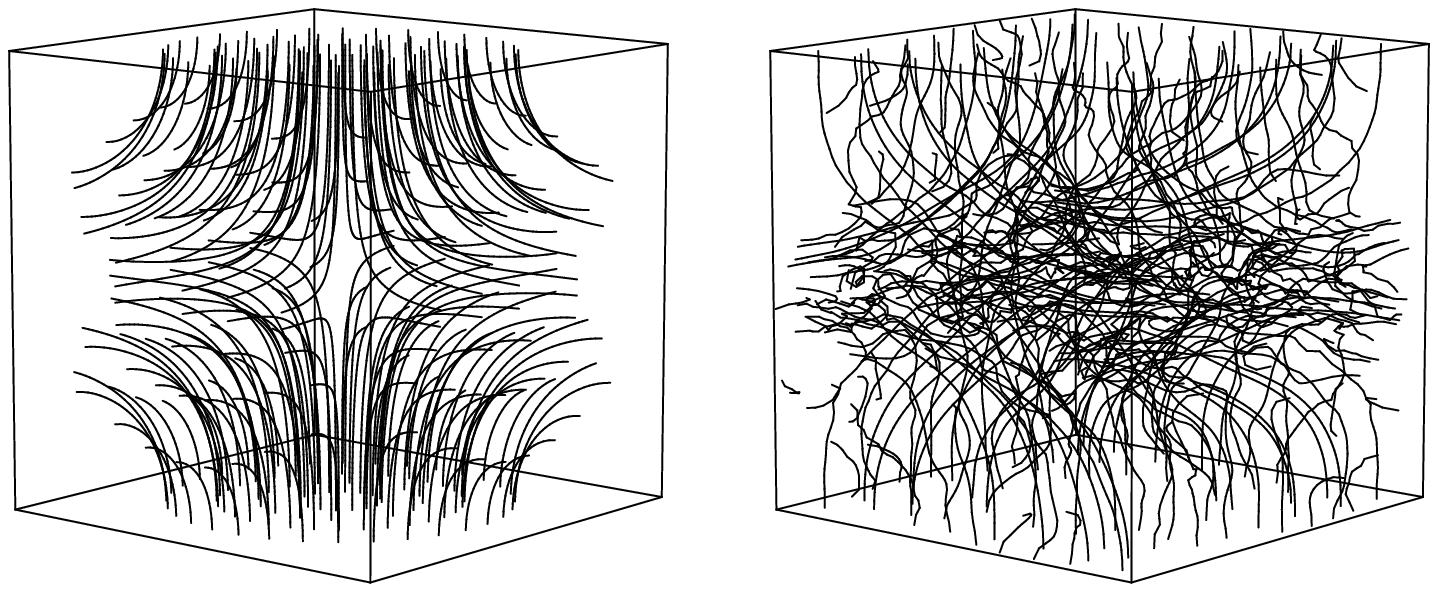}
 (a) \hspace{3cm} (b)
\end{center}
\end{minipage}

\begin{minipage}{1.0\linewidth}
\begin{center}
\epsfxsize=8cm \epsfbox{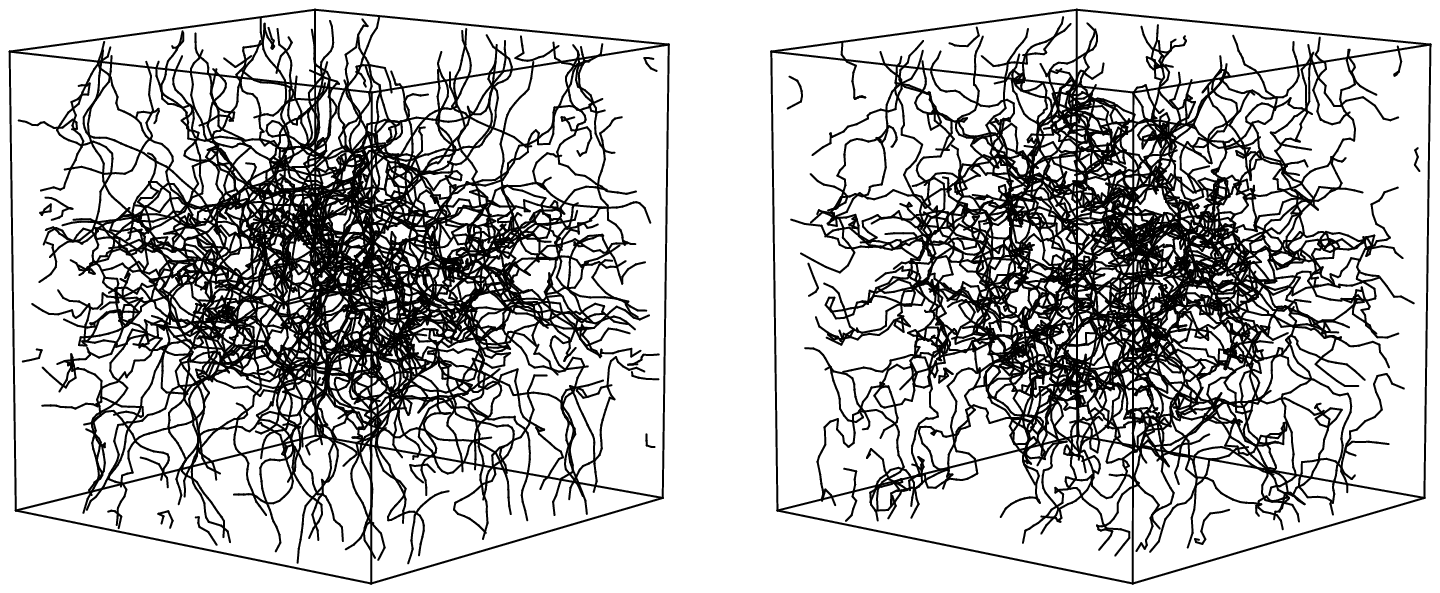}
 (c) \hspace{3cm} (d)
\end{center}
\end{minipage}
\caption{Time evolution of the vortex tangle at $t$=0 sec(a), $t$=30.0 sec(b),
$t$=50.0 sec(c) and $t$=70.0 sec(d).}
 \label{eps1}
\end{figure}

First we discuss the transient behavior of the $k$ dependence of the energy spectrum
$E(k)$. The energy spectra calculated from each configuration in Fig. 1 are shown in
Fig. 2. It is shown that the slope is changed about at $k=2\pi/l$, where $l$ is the
average intervortex spacing. The energy spectrum for $k>2\pi/l$ has $k^{-1}$
behavior which comes from the velocity field near each vortex line \cite{Nore,1/k},
though the random vortices compose the turbulent velocity field. On the contrary,
the spectrum for $k<2\pi/l$ is strongly affected by the random vortex configuration.
At $t=0$ sec, the spectrum has a large peak at the smallest wave number, being flat
in the intermediate range because there are only large vortices and no short-scale
structure on them. Figure 2 shows that as the vortices become the homogeneous and
isotropic vortex tangle the slope for $k<2\pi/l$ converges to the Kolmogorov form
$k^{-5/3}$.

\begin{figure}[tbhp]
\begin{center}
\epsfxsize=8cm \epsfbox{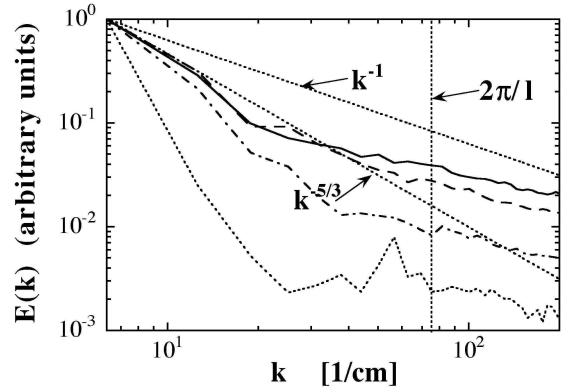}
\end{center}
\caption{The energy spectra of the vortex tangle at $t$=0 sec (dashed), $t$=30.0
sec (dot-dashed), $t$=50.0 sec (long-dashed) and $t$=70.0 sec (solid).}
 \label{eps2}
\end{figure}

The Kolmogorov law can be derived from the argument based on the picture of the
cascade process\cite{Frish}. In the inertial range the kinetic energy is transferred
steadily from small $k$ to large $k$ without dissipation, and dissipated at the end
of the inertial range. Thus, for the steady state, the energy dissipation rate
$\epsilon=-dE/dt$ can be identified with the energy flux in the inertial range. Then
the energy spectrum depends only on the wave number $k$ and the energy dissipation
rate $\epsilon=-dE/dt$, which leads to the Kolmogorov spectrum
$E(k)=C\epsilon^{2/3}k^{-5/3}$. Here $C$ is the (dimensionless) Kolmogorov constant
of order unity.

In our previous papers, the cascade process without the mutual friction in
superfluid turbulence was discussed\cite{Tsubota}. Through lots of reconnections,
the vortex tangle breaks up to smaller ones and this process proceeds
self-similarly, and in our calculation the smallest vortex whose size is the order
of the numerical space resolution $\Delta \xi$ is eliminated by the cutoff
procedure. This resolution in our calculation is much larger than the dissipative
scale in real system. However this numerical cutoff can be justified, because the
cascade process at a small scale proceeds much faster than that at a large scale.
Actually we showed the decay rate of the density of vortices was almost independent
of the cutoff scale $\Delta \xi$. Figure 3 shows the energy dissipation rate
$\epsilon$ due to the cutoff procedure in the dynamics of Fig. 1. After 70 sec, the tangle becomes isotropic and homogeneous losing the memory of the initial configuration, so the change of the dissipation rate becomes slow free from the artifact of the early large dissipation.

\begin{figure}[tbhp]
\begin{center}
\epsfxsize=8cm \epsfbox{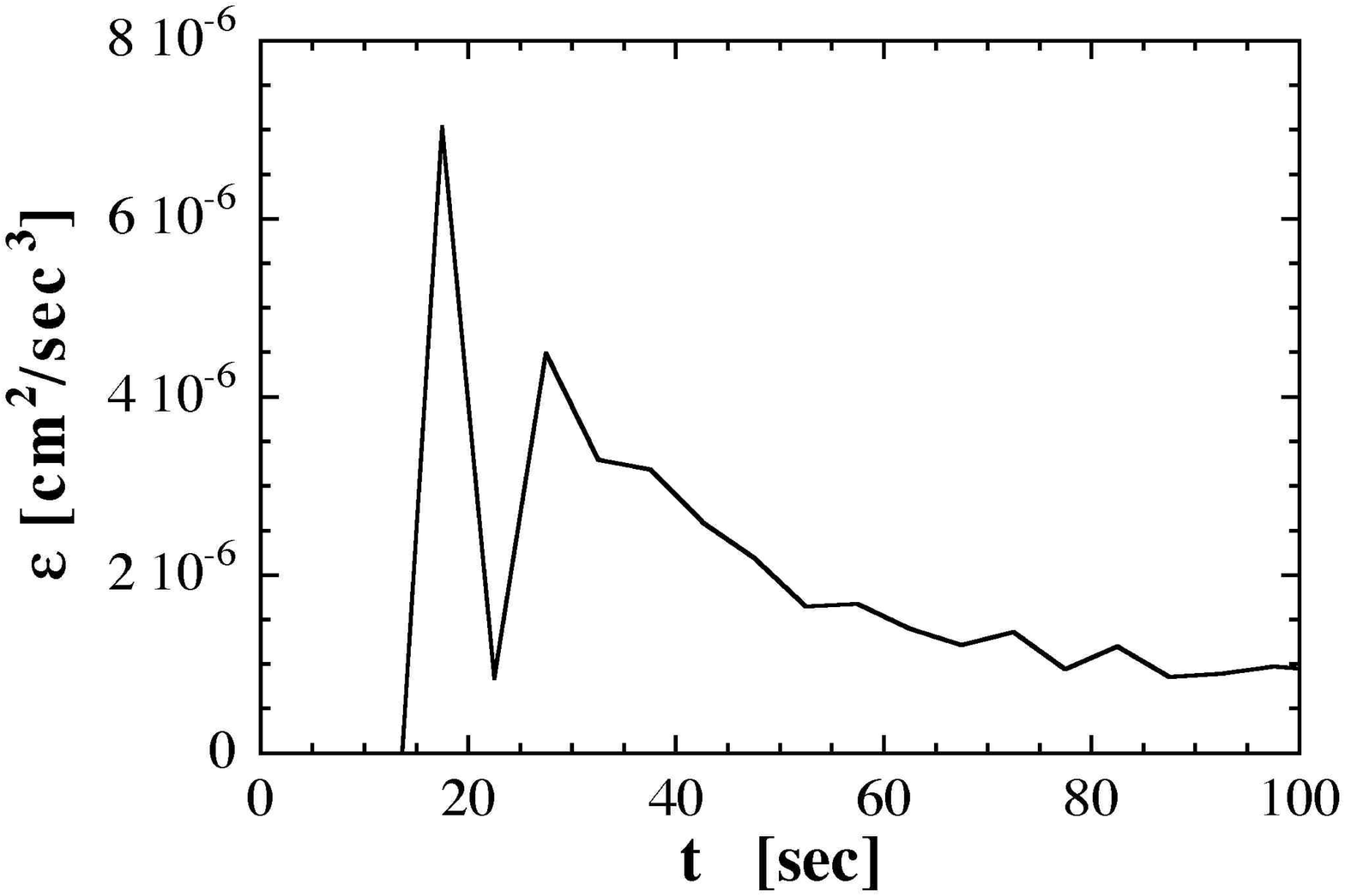}
\end{center}
\caption{The energy dissipation rate $\epsilon$ which calculated from the dynamics
of Fig. 1.}
 \label{eps3}
\end{figure}

Next we compare quantitatively the energy spectrum at 70 sec with the
Kolmogorov law $E(k)=C\epsilon^{2/3}k^{-5/3}$. The Kolmogorov constant $C$ is known
as the parameter of order unity. Here we use $C=1$ and $\epsilon=1.287
\times 10^{-6}$ cm$^2$/sec$^3$. Then we can determine uniquely the energy
spectrum. Figure 4 shows the comparison between the energy spectrum at $t=70$ sec
and the Kolmogorov law with $C=1$ and $\epsilon=1.287 \times 10^{-6}$ cm$^2$/sec$^3$. 
The energy spectrum for $k<2\pi/l$ is consistent with the
Kolmogorov law not only on the wave number dependence but also on the absolute
value. The dissipative mechanism due to the cutoff works only at the largest wave
number $k \sim 2 \pi/\Delta \xi=343$ cm$^{-1}$ . However the energy spectrum at
small $k$ region is determined by that dissipation rate. This result supports just
the picture of the inertial range. Although our spectrum has $k^{-1}$ region between
the Kolmogorov region and the dissipative wave number, the energy flux exists also
in this region. The spectrum in this region includes the contribution coming from
each vortex line and that of the energy flux, while the former is dominant
\cite{1/k}. The cascade process in this region will be discussed elsewhere.

The Kolmogorov law is the scaling property in $k$ space, and closely related with
the self-similarity of the turbulent velocity field in the real space. We devote the
rest part of this paper to the following question: is this scaling property in $k$
space related with the self-similarity of the vortex tangle in real space or not?
However, in conventional turbulence it is very difficult to discuss this problem,
because the viscous diffusion of vorticity makes the vortex configuration obscure.
On the contrary, the characters of the superfluid turbulence are the definiteness of
the vortex line due to the absence of the viscosity and the fixed circulation by
the quantum effect. These characters allow us to describe the system by the
topological configuration of a vortex tangle. Hence, in order to discuss the
self-similarity in a real space, it is meaningful to investigate a vortex length
distribution (VLD) $n(x)$, where $n(x) \Delta x$ represents the number of the
vortices whose length is from $x$ to $x+\Delta x$.

\begin{figure}[tbhp]
\begin{center}
\epsfxsize=8cm \epsfbox{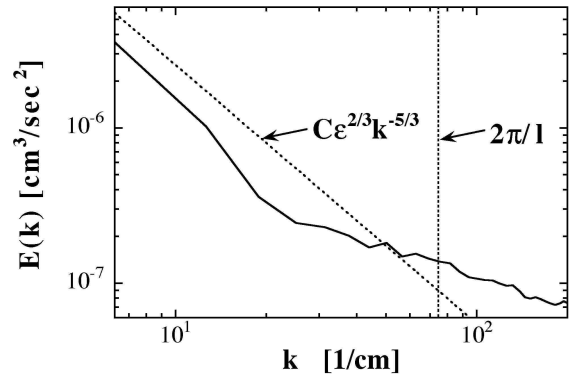}
\end{center}
\caption{Comparison of the energy spectrum (solid line) at $t=70$ sec with the
Kolmogorov law $E(k)=C\epsilon^{2/3}k^{-5/3}$ (dotted line) with $C=1$ and
$\epsilon=1.287 \times 10^{-6}$ cm$^2$/sec$^3$.}
 \label{eps4}
\end{figure}

In order to suppress the fluctuation of the VLD, we use the time averaging procedure
in the interval 4.0 sec. Figure 5 shows the averaged VLD at each time, where the
largest scale in $x$ axis is the size of the cube and the smallest scale is the
length of the numerical space resolution. At large scale the VLD is strongly affected by the effect of the boundary, and at small scale decreases rapidly by the effect of the cutoff procedure. As the vortices approach the homogeneous
and isotropic tangle, in the intermediate range the VLD becomes to obey a scaling property $n(x) \propto x^{\alpha}$. Using the least squares fits, we determine this scaling exponent $\alpha$ at $t=60.0$ sec in the region of $0.15 < x < 0.65$ [cm], and find $\alpha = -1.34 \pm 0.18$.

In conclusion we have studied numerically the energy spectrum of the superfluid
turbulence without the mutual friction. For $k>2\pi/l$, the spectrum can be
attributed to the contribution of each vortex line. For $k<2\pi/l$, as the vortices approach the homogeneous and isotropic tangle, the slope of the spectrum converges to the Kolmogorov form $k^{-5/3}$. By using the energy dissipation
 rate due to the elimination of the smallest vortices, the spectrum for $k<2\pi/l$
 is consistent with the Kolmogorov law not only on the wave number dependence but also
 on the absolute value. The VLD $n(x)$ becomes to obey a scaling property reflecting the self-similarity of the tangle.

\begin{figure}[tbhp]
\begin{center}
\epsfxsize=8cm \epsfbox{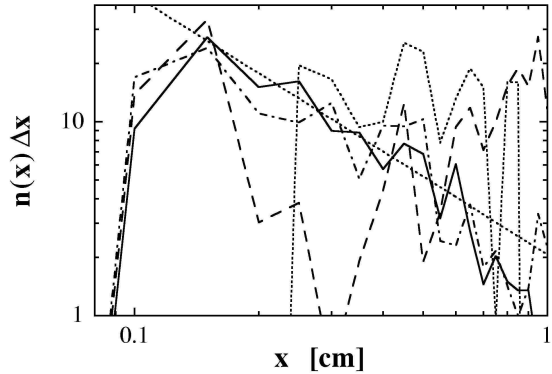}
\end{center}
\caption{The VLD $n(x)\Delta x$ at $t$=0 sec (dashed), $t$=20.0 sec (long-dashed),
$t$=40.0 sec (dot-dashed) and $t$=60.0 sec (solid) for $\Delta x$=0.05. The dashed line is determined by the least squares fits at $t=60.0$ sec.}
 \label{eps5}
\end{figure}

MT acknowledges support by a Grant-in-Aid for Scientific Research (Grant No. 12640357) by Japan Society for the Promotion of Science. 
SN thanks Russian Foundation of Basic Research (Grant N 99-02-16942) for supporting that field.

%


\begin{thebibliography}{20}
\bibitem{1/k} The energy spectrum of a straight vortex line can be derived. Since
the velocity field of a straight vortex line is $\kappa/2\pi r$, the kinetic energy
of a vortex line per unit mass is calculated as $E=(1/2) \int v^2 (\Vec{r}) d^3
r=(\kappa^2 L / 4\pi) \ln (D/a)$, where $L$ is the total line length per unit
volume and $r$ is the distance from the vortex line. Since the energy spectrum
$E(k)=Ck^\alpha$ can be defined as $E=\int_{2\pi/D}^{2\pi/a} E(k)dk$, the parameters
$\alpha$ and $C$ are determined as $\alpha=-1$ and $C=\kappa^2 L / 4\pi$. For
example, at $t=$70 sec, $L=151.9$ cm$^{-2}$, so that $E(k)_{\rm line}=(\kappa^2 L / 4\pi)
 k^{-1}=1.20 \times 10^{-7}$ cm$^3$/sec$^2$ for $k=$100 cm$^{-1}$. This is
consistent with our numerical result $E(k)=1.07 \times 10^{-7}$ cm$^3$/sec$^2$ for
$k=$100 cm$^{-1}$ in Fig. 4.
\end{thebibliography}
\end{document}